\documentclass[12pt]{article}
\usepackage{graphics}

\def\<~{\stackrel{<}{\mbox{\scriptsize $\sim$}}}
\def\>~{\stackrel{>}{\mbox{\scriptsize $\sim$}}}
\def\eq#1{(\ref{#1})}

\begin{document}

\title{
Quintessence and the Transition to an Accelerating Universe}
\author{
Carl L. Gardner\\
{\em gardner@math.asu.edu}\\
Department of Mathematics and Statistics\\ 
Arizona State University\\
Tempe AZ 85287-1804
}
\date{}

\maketitle
\thispagestyle{empty}

\begin{abstract}

The implications of seven popular models of quintessence based on
supergravity or M/string theory for the transition from a decelerating
to an accelerating universe are explored.

All seven potentials can mimic the $\Lambda$CDM model at low redshifts
$0 \le z \le 5$.  However, for a natural range of initial values of
the quintessence field, the SUGRA and Pol\'onyi potentials predict a
transition redshift $z_t \approx 0.5$ for $\Omega_{\Lambda0} = 0.70$,
in agreement with the observational value $z_t \approx 0.46$ and in
mild conflict with the $\Lambda$CDM value $z_t = 0.67$.

Tables are given for the quintessence potentials for the recent
average $\overline{w}_0$ of the equation of state parameter, and for
$w_0$ and $w_1$ in the low-$z$ approximation $w \approx w_0 + w_1 z$.

It is argued that for the exponential potential $e^{\lambda \varphi}$
to produce a viable present-day cosmology, $\lambda \le \sqrt{2}$.

A robust, scaled numerical method is presented for simulating the
cosmological evolution of the scalar field.

\end{abstract}

\section{Introduction}

In the standard $\Lambda$CDM cosmological model, the universe makes a
transition from deceleration to acceleration at a redshift $z_t =
0.67$ for $\Omega_{\Lambda0} = 0.70$.  This prediction is to be
contrasted with the observational value $z_t = 0.46 \pm 0.13$ from the
distance-redshift diagram for type Ia supernovae (SNe
Ia)~\cite{Riess:2004nr}.  With further observations, the SNe Ia data
may converge to the $\Lambda$CDM value.  However the value $z_t
\approx 0.46$ could be a signal of the effects of a quintessence
scalar field, extra spatial dimensions, and/or modifications to
general relativity.

For the spatially homogeneous quintessence scalar field $\phi$, define
the equation of state parameter $w = w(z) = P_\phi/\rho_\phi$, where
the scalar field pressure $P_\phi$ and energy density $\rho_\phi$ are
given by
\begin{equation}
	P_\phi = \frac{1}{2} \dot{\phi}^2 - V(\phi) ,~~
	\rho_\phi = \frac{1}{2} \dot{\phi}^2 + V(\phi) .
\end{equation}
The SNe Ia observations~\cite{Riess:2004nr} bound the recent ($z \le
1.75$) average $\overline{w}_0 < -0.76$ (95\% CL) assuming
$\overline{w}_0 \ge -1$, and measure $z_t = 0.46 \pm 0.13$.
Alternatively, the SNe Ia data place bounds $-1 < w_0 < -0.72$ (95\%
CL) and $w_1 = 0.6 \pm 0.5$, where $w(z) \approx w_0 + w_1 z$ for $z
\<~ 1$.  This investigation will explore seven popular models of
quintessence (see Table~\ref{potentials}), and compare and contrast
their predictions for $z_t$, $\overline{w}_0$, $w_0$, and $w_1$.
These models are basically the ones analyzed in Ref.~\cite{KL2} in
terms of dark energy and the fate of the universe (see also
Refs.~\cite{KL1,KL3}).

We will assume a flat Friedmann-Robertson-Walker universe.  In the
$\Lambda$CDM model, the total energy density $\rho = \rho_m + \rho_r +
\rho_\Lambda = \rho_c$, where $\rho_c$ is the critical density for a
flat universe and $\rho_m$, $\rho_r$, and $\rho_\Lambda$ are the
energy densities in (nonrelativistic) matter, radiation, and the
cosmological constant respectively.  In the quintessence/cold dark
matter (QCDM) model, $\rho = \rho_m + \rho_r + \rho_\phi = \rho_c$.
Ratios of energy densities to the critical density will be denoted by
$\Omega_m = \rho_m/\rho_c$, $\Omega_r = \rho_r/\rho_c$,
$\Omega_\Lambda = \rho_\Lambda/\rho_c$, and $\Omega_\phi =
\rho_\phi/\rho_c$, while ratios of present energy densities to the
present critical density will be denoted by $\Omega_{m0}$,
$\Omega_{r0}$, $\Omega_{\Lambda0}$, and $\Omega_{\phi0} \equiv
\Omega_{\Lambda0}$.

\begin{table}[ht]
\center{
\begin{tabular}{|l l|} \hline 
{\em dimensionless}\/ $V$ & {\em name} \\ \hline
$e^{\lambda \varphi}$ & exponential \\
$\cosh(\sqrt{2} \varphi)$ & cosh (stable de Sitter)\\
$2 - \cosh(\sqrt{2} \varphi)$ & cosh (unstable de Sitter)\\
$1 + \cos(\varphi)$ & axion \\
$\cos(\varphi)$ & axion (unstable de Sitter)\\
$\left[ \left( 1 + \frac{\varphi}{\sqrt{2}} \left(\frac{\varphi}{\sqrt{2}} + 
	\beta\right)\right)^2 - 3 \left(\frac{\varphi}{\sqrt{2}} + 
	\beta\right)^2 \right] e^{\varphi^2/2}$ & 
Pol\'onyi \\
$e^{\varphi^2/2}/\varphi^4$ & SUGRA \\
\hline
\end{tabular}
}
\caption{Quintessence potentials inspired by supergravity or M/string theory.
$\varphi = \phi/M_P$.}
\label{potentials}
\end{table}

In the $\Lambda$CDM model,
\begin{equation}
	z_t = \left( \frac{2 \Omega_{\Lambda0}}{\Omega_{m0}} \right)^{1/3}
	- 1 .
\end{equation}
From WMAP $+$ SDSS~\cite{Tegmark:2003ud}, $\Omega_{\Lambda0}$ =
0.71$^{+0.03}_{-0.05}$.  For the $1 \sigma$ lower bound on
$\Omega_{\Lambda0}$, $z_t$ = 0.57, which is just at the upper $1
\sigma$ bound for the measured $z_t$.  Thus the $\Lambda$CDM model
value for $z_t$ lies at the boundary of the joint 68\% confidence
interval of the SNe Ia data.  We are here interested, however, in
whether quintessence models satisfying the observational bounds on
$\overline{w}_0$ and $w_0$ may be in better agreement with the
measured central value for $z_t$ and consistent with the 1$\sigma$
limits on $w_1$.  Of the seven models in Table~\ref{potentials}, all
but two are very close to the $\Lambda$CDM model values for $z_t$ and
$w_1 = 0$ (in fact, $z_t \ge 0.67$ for $\Omega_{\phi0}$ = 0.70), while
the SUGRA and Pol\'onyi potentials differ qualitatively from the
others in their predictions for $z_t$ and $w_1$, and in a certain
natural parameter range agree closely with the observed central
values.

All seven potentials can mimic the $\Lambda$CDM model at low redshifts
$0 \le z \le 5$ to well within the observational error bounds.  If the
SNe Ia data converge to the $\Lambda$CDM value for $z_t$, then further
restrictions can be placed on the possible initial values for $\phi$
and on parameters in the potentials.  For the SUGRA and Pol\'onyi
potentials to mimic a cosmological constant at present, the initial
values for $\phi$ must be fine tuned; these models can naturally
predict $z_t \approx 0.5$ for $\Omega_{\Lambda0} = 0.70$.

Ref.~\cite{Brax:2001ah} gave $w_0$ equal to $-0.8$ to $-0.9$ and
$w_1 \approx$ 0.3--0.45 for the SUGRA potential
$e^{\varphi^2/2}/\varphi^\alpha$ for $2 \le \alpha \le 16$---in
agreement with our results for $\alpha = 4$ for a range of
initial values $0 < \varphi_i = \phi_i/M_P \<~ 1$.

Curves for $\Omega_\phi$ and $w(z)$ for the cosh, SUGRA ($\varphi_i =
1$ and $\alpha$ = 11 only), and Pol\'onyi ($\varphi_i = -1$ only)
potentials were given in Ref.~\cite{KL2}; where they overlap, our
results agree with theirs.  The main focus here, however, is on the
transition redshift, which was not addressed in Ref.~\cite{KL2}.

Mention should also be made of Ref.~\cite{A-P}, which investigated the
implication of a 5D brane world modification of general
relativity~\cite{DGP} for the recent acceleration of the universe and
found $z_t \approx 0.5$.

\section{Cosmological Equations}

The homogeneous scalar field obeys the Klein-Gordon equation
\begin{equation}
	\ddot{\phi} + 3 H \dot{\phi} = -\frac{d V}{d \phi} \equiv - V_\phi .
\label{phi}
\end{equation}
The Hubble parameter $H$ is related to the scale factor $a$ and the
energy densities in matter, radiation, and the quintessence field
through the Friedmann equation
\begin{equation}
	H^2 = \left( \frac{\dot{a}}{a} \right)^2 = \frac{\rho}{3 M_P^2} =
	\frac{1}{3 M_P^2} \left( \frac{1}{2} \dot{\phi}^2 + V(\phi) 
	+ \rho_m + \rho_r \right)
\label{H}
\end{equation}
where the (reduced) Planck mass $M_P = 2.4 \times 10^{18}$ GeV.

We will use the logarithmic time variable $\tau = \ln(a/a_0) =
-\ln(1+z)$.  Note that for de Sitter space $\tau = H_\Lambda t$, where
$H_\Lambda^2 = \rho_\Lambda/(3 M_P^2)$, and that $H_\Lambda t$ is a
useful time variable for the era of $\Lambda$-matter domination (see
e.g.\ Ref.~\cite{alpha}).  Also note that for $0 \le z \le z_{BBN} =
10^{10}$, $-23.03 \le \tau \le 0$, where BBN denotes the era of
big-bang nucleosynthesis.  (BBN occurs over a range of $z \approx
10^9$--$10^{10}$; we will take $z_{BBN} \equiv 10^{10}$.)

For numerical simulations, the cosmological equations should be put
into dimensionless form.  Eqs.~\eq{phi} and~\eq{H} can be cast in the
form of a system of two first-order equations in $\tau$ plus a scaled
version of $H$:
\begin{equation}
	\overline{H} \varphi' = \psi
\label{phi-1}
\end{equation}
\begin{equation}
	\overline{H} (\psi' + 3 \psi) = - 3 \overline{V}_\varphi
\label{phi-2}
\end{equation}
\begin{equation}
	\overline{H}^2 = \frac{1}{6} \psi^2 + \overline{V} + 
	\overline{\rho}_m + \overline{\rho}_r
\label{scaled-H}
\end{equation}
where $\overline{H} = H/H_0$, $\varphi = \phi/M_P$, $\overline{V} =
V/\rho_{c0}$, $\overline{\rho}_m = \rho_m/\rho_{c0}$,
$\overline{\rho}_r = \rho_r/\rho_{c0}$, and a prime denotes
differentiation with respect to $\tau$: $\varphi' = d \varphi/d \tau$,
etc.

A further scaling may be performed resulting in a set of equations
which is numerically more robust, especially before the time of BBN:
\begin{equation}
	\tilde{H} \varphi' = \tilde{\psi}
\label{phi-tilde-1}
\end{equation}
\begin{equation}
	\tilde{H} (\tilde{\psi}' + \tilde{\psi}) = 
	- 3 \tilde{V}_\varphi
\label{phi-tilde-2}
\end{equation}
\begin{equation}
	\tilde{H}^2 = \frac{1}{6} \tilde{\psi}^2 + \tilde{V} + 
	\tilde{\rho}_m + \tilde{\rho}_r
\label{scaled-H-tilde}
\end{equation}
where $\tilde{H} = e^{2 \tau} H/H_0$, $\tilde{\psi} = e^{2 \tau}
\psi$, $\tilde{V} = e^{4 \tau} V/\rho_{c0}$, $\tilde{\rho}_m = e^{4
\tau} \rho_m/\rho_{c0}$, and $\tilde{\rho}_r = e^{4 \tau}
\rho_r/\rho_{c0}$.

\begin{figure}[htbp]
\center{
\scalebox{1.1}{\includegraphics{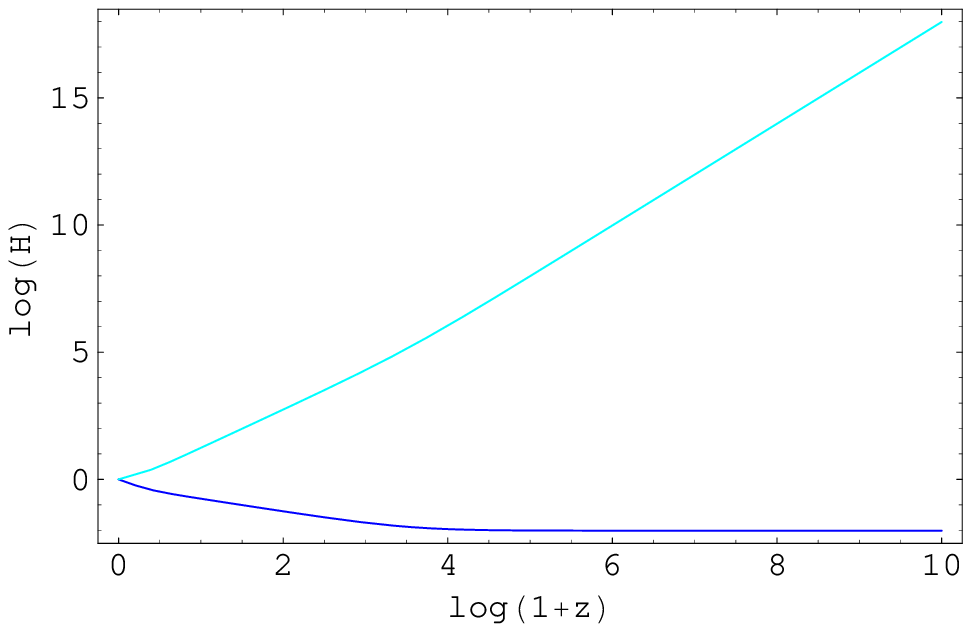}}
}
\caption{Log$_{10}$ of $\tilde{H}$ (blue, bottom)
vs.\ $\overline{H}$ (cyan, top) for the $e^{\sqrt{2} \varphi}$ potential.}
\label{fig-H}
\end{figure}

Figure~\ref{fig-H} illustrates (for the exponential potential
$e^{\sqrt{2} \varphi}$) that while $\tilde{H}$ varies over only two
orders of magnitude between BBN and the present, $\overline{H}$ varies
over eighteen orders of magnitude.  A similar scaling effect occurs for
$\tilde{\psi}$ vs.\ $\psi$.

We define the recent average of the equation of state parameter $w$ by
rewriting the conservation of energy equation
\begin{equation}
	\dot{\rho} + 3 H (\rho + P) = 0
\label{energy}
\end{equation}
where $P$ is the pressure, as
\begin{equation}
	0 = \rho' + 3 (\rho + P) = \sum_j \rho'_j +
	3 \sum_j (1 + w_j) \rho_j
\label{energy'}
\end{equation}
where $j$ = $m$, $r$, $\phi$.  The solution is
\begin{equation}
	\rho_j = \rho_{j0} \exp\left\{-3 \int_0^\tau (1 + w_j) d \tau 
	\right\} .
\label{rhoj}
\end{equation}
Note that $w_m = 0$ and $w_r = 1/3$ are constant except near
particle-antiparticle thresholds.  The recent average of $w_\phi
\equiv w$ is defined as
\begin{equation}
	\overline{w}_{0} = \frac{1}{\tau} \int_0^\tau w d \tau .
\label{wbar}
\end{equation}
We will take the upper limit of integration $\tau$ to correspond to $z
= 1.75$.


Strictly speaking, 
\begin{equation}
	\rho_m = \rho_{m0} e^{-3 \tau} f_m(\tau) ,~~
	\rho_r = \rho_{r0} e^{-4 \tau} f_r(\tau)
\label{f}
\end{equation}
where $f_m(\tau)$ and $f_r(\tau)$ (with $f_m(0) = 1 = f_r(0)$) account
for the change in the effective number ${\cal N}(T)$ of massless
degrees of freedom as $\tau$ decreases and the temperature $T$ of the
gas of relativistic particles increases.  Below $T$ = 1 MeV at
$z_{BBN}$, ${\cal N}$ = 3.36 is constant, so we can safely set
$f_m(\tau) = 1$ since $\rho_m \sim (1+z)^3$ quickly becomes negligible
compared to $\rho_r \sim (1+z)^4$ for $z > z_{m-r}$ = 3233 at the
equality of matter and radiation densities.  In computing the
evolution of the quintessence field, we will start with initial
conditions at $z_{BBN}$, so we can also set $f_r(\tau) = 1$ for our
purposes.  Thus in Eqs.~\eq{scaled-H} and~\eq{scaled-H-tilde}, 
\begin{equation}
	\overline{\rho}_m + \overline{\rho}_r =
	\Omega_{m0} e^{-3 \tau} + \Omega_{r0} e^{-4 \tau}
\label{scaled-rho-1}
\end{equation}
\begin{equation}
	\tilde{\rho}_m + \tilde{\rho}_r =
	\Omega_{m0} e^\tau + \Omega_{r0} .
\label{scaled-rho-2}
\end{equation}
(Ref.~\cite{A-G} suggests the phenomenological form $f_r(\tau) =
e^{-\tau/15}$ for $z$ going as far back as $10^{30}$.)

The transition redshift $z_t$ is defined through the acceleration
Friedmann equation
\begin{equation}
	\frac{\ddot{a}}{a} = - \frac{1}{6 M_P^2} ( \rho + 3 P )
\label{acc}
\end{equation}
which may be written in the form
\begin{equation}
	-q = \frac{1}{H^2} \frac{\ddot{a}}{a} = \frac{H'}{H} = 
	- \frac{1}{2} ( \Omega_m + 2 \Omega_r + (1 + 3 w) \Omega_\phi )
\label{q}
\end{equation}
where $-q$ is the acceleration parameter.  The Friedmann
equation~\eq{H}, conservation of energy equation~\eq{energy}, and the
acceleration equation~\eq{acc} are related by the Bianchi identities,
so that only two are independent.  Eq.~\eq{energy} gives the
evolution~\eq{f} of $\rho_m$ and $\rho_r$, and the Klein-Gordon
equation~\eq{phi} for the weakly coupled scalar field.  When a
cosmological model involves a collapsing stage where $H$ reverses
sign, Eq.~\eq{acc} should be used instead of Eq.~\eq{H}.  In
computational form, the acceleration equation becomes
\begin{equation}
	\overline{H} ~ \overline{H}' = 
	- \frac{1}{2} \overline{\rho}_m - \overline{\rho}_r - 
	\frac{1}{3} \psi^2 + \overline{V}
\label{scaled-acc}
\end{equation}
or
\begin{equation}
	\tilde{H} \tilde{H}' - 2 \tilde{H}^2 = 
	- \frac{1}{2} \tilde{\rho}_m - \tilde{\rho}_r - 
	\frac{1}{3} \tilde{\psi}^2 + \tilde{V} .
\label{scaled-acc-tilde}
\end{equation}


\section{Simulations}

For the computations below, we will use
Eqs.~\eq{phi-tilde-1}--\eq{scaled-H-tilde} with initial conditions
specified at $z_{BBN}$ by $\varphi_i$ and $\dot{\varphi}_i \propto
\psi_i = 0$.  The potential $V = A \cdot$(dimensionless potential),
where the dimensionless potentials are given in
Table~\ref{potentials}.  The constant $A$ is adjusted by a bisection
search method so that $\Omega_{\phi0} = \Omega_{\Lambda0}$.  This
involves the usual single fine tuning.

Since several observational lines including SNe Ia, the cosmic
microwave background (CMB), large scale structure (LSS) formation, the
integrated Sachs-Wolfe effect, and gravitational lensing measure $0.66
\le \Omega_{\Lambda0} \le 0.74$, we will restrict our analysis to
this interval, even though technically the bounds are 1$\sigma$.  Our
main line of development will take $\Omega_{\Lambda0} = 0.70$; in
passing, we will make some remarks about what changes if
$\Omega_{\Lambda0}$ = 0.66.  The main effect of changing
$\Omega_{\Lambda0}$ to 0.66 (0.74) is to shift the acceleration curves
toward the left (right).

We will consider an ultra-light scalar field with $m_\phi^2 \sim
H_\Lambda^2$; then $\phi$ ``sits and waits'' during the early
evolution of the universe, and only starts moving when $H^2 \sim
m_\phi^2$.  In this way it is easy to satisfy the BBN ($z \sim
10^9$--$10^{11}$), CMB ($z \sim 10^3$--$10^5$), and LSS ($z \sim
10$--$10^4$) bounds on $\Omega_\phi \<~ 0.1$.  An ultra-light scalar
field also reflects the observational evidence that the universe has
only recently made the transition from deceleration to acceleration
and has only recently become dominated by dark energy.

Ultra-light scalar fields exist near de Sitter space extrema in 4D
extended gauged supergravity theories (with noncompact internal
spaces), with quantized mass squared~
\cite{Gates,Hull,KL1,Gibbon,Fre,Kallosh}
\begin{equation}
	m^2 = n H_\Lambda^2 ,~~ 
	H_\Lambda^2 =  \frac{\rho_{\Lambda Q}}{3 M_P^2}
\label{m}
\end{equation}
where $-6 \le n \le 12$ is an integer.  In this context $H_\Lambda$ is
the de Sitter space value of $H$ with effective cosmological constant
$\rho_{\Lambda Q}$ at the extremum of the quintessence potential $V$.
Note that to produce the current acceleration of the universe,
typically $\rho_{\Lambda Q} \approx \rho_\Lambda$, but does not equal
$\rho_\Lambda$ unless the quintessence field---unlike the ones
below---is at a de Sitter extremum at $t_0$.  In certain cases, these
theories are directly related to M/string theory.  An additional
advantage of these theories is that the classical values $m^2 = n
H_\Lambda^2$ and $\rho_{\Lambda Q}$ are protected against quantum
corrections.  The relation $m^2 = n H_\Lambda^2$ was derived for
supergravity with scalar fields; in the presence of other matter
fields, the relation may be modified.

\subsection{Exponential Potential}

The exponential potential $e^{\lambda
\varphi}$~\cite{Wetterich:1987fm,Ferreira:1997hj,Copeland:1997et,Doran:2002ec}
can be derived from M-theory~\cite{Townsend:2001ea} or from $N$ = 2,
4D gauged supergravity~\cite{Andrianopoli:1996cm}.  The results for $V =
A e^{\lambda \varphi}$ are independent of the initial value
$\varphi_i$, which we arbitrarily set equal to 1.

\begin{table}[htb]
\center{
\begin{tabular}{|l l l l l|} \hline 
$\lambda$ & $\overline{w}_0$ & $z_t$ & $w_0$ & $w_1$ \\ \hline
$1/\sqrt{3}$ & $-0.98$ & 0.68 & $-0.95$ & $-0.07$ \\ $1$ & $-0.93$ &
0.71 & $-0.84$ & $-0.19$ \\ $\sqrt{2}$ & $-0.83$ & 0.76 & $-0.68$ &
$-0.33$ \\ $\sqrt{3}$ & $-0.70$ & 0.76 & $-0.49$ & $-0.40$ \\ $2$ &
$-0.50$ & & $-0.27$ & $-0.37$ \\
\hline
\end{tabular}
}
\caption{Parameters for the potential $e^{\lambda \varphi}$.}
\label{exponential}
\end{table}

\begin{figure}[htbp]
\center{
\scalebox{1.1}{\includegraphics{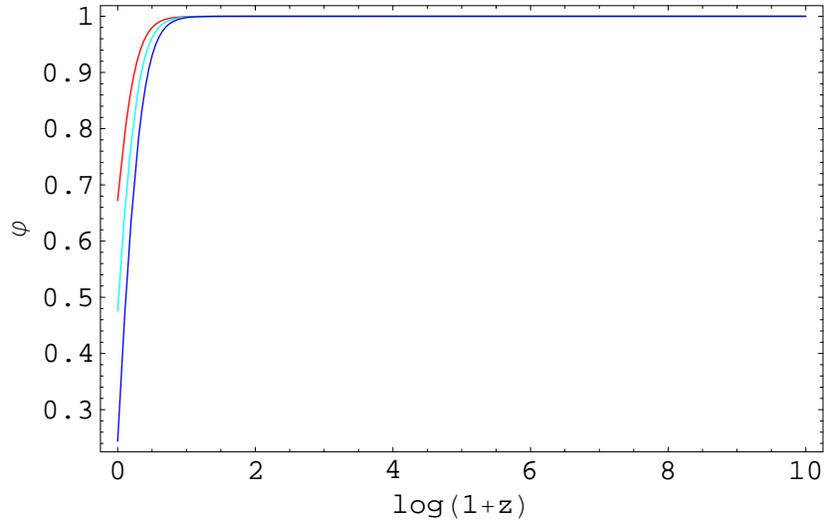}}
}
\caption{$\varphi(\tau)$ for the exponential potential $e^{\lambda \varphi}$.
$\lambda$ = 1 (red), $\sqrt{2}$ (cyan), and $\sqrt{3}$ (blue)
from top to bottom.}
\label{fig-exp-phi}
\end{figure}

\begin{figure}[htbp]
\center{
\scalebox{1.1}{\includegraphics{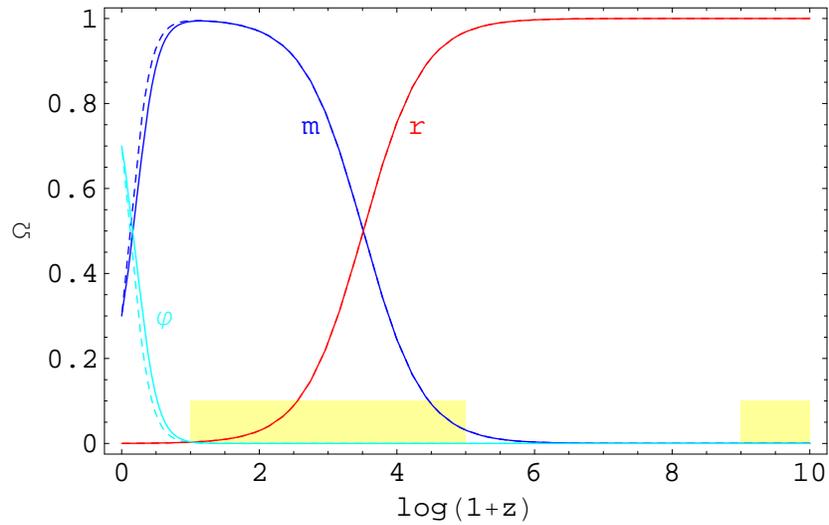}}
}
\caption{$\Omega$ for the exponential potential $e^{\sqrt{2} \varphi}$
(solid) vs.\ $\Lambda$CDM (dotted).  The light yellow rectangles are
the bounds on $\Omega_\phi$ from LSS, CMB, and BBN.}
\label{fig-exp-Omega}
\end{figure}

\begin{figure}[htbp]
\center{
\scalebox{1.1}{\includegraphics{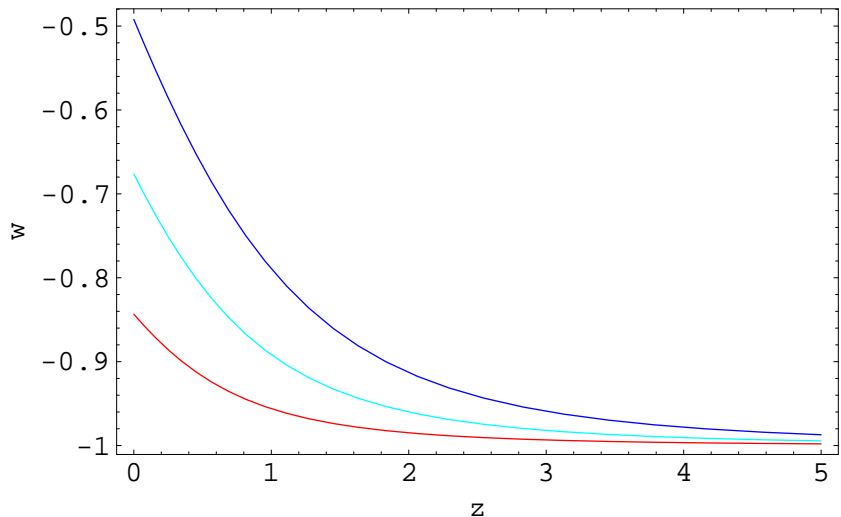}}
}
\caption{$w$ for the exponential potential $e^{\lambda \varphi}$.
$\lambda$ = 1 (red), $\sqrt{2}$ (cyan), and $\sqrt{3}$ (blue)
from bottom to top.}
\label{fig-exp-w}
\end{figure}

\begin{figure}[htbp]
\center{
\scalebox{1.1}{\includegraphics{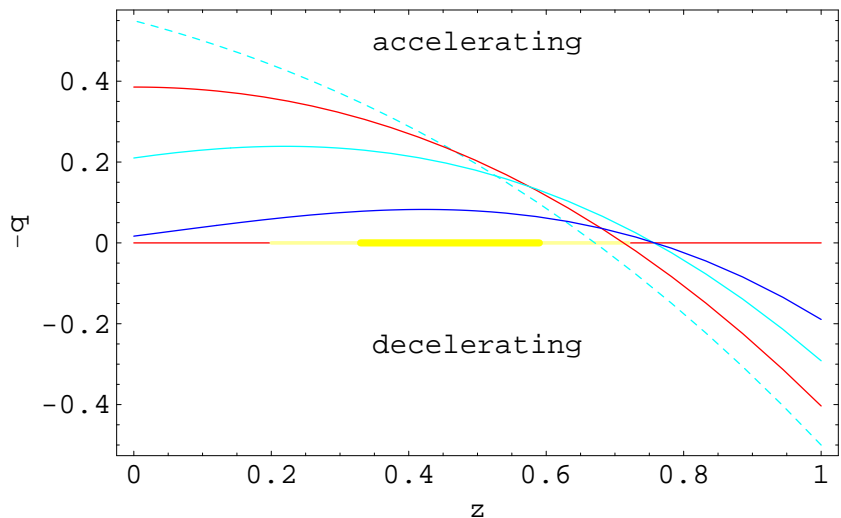}}
}
\caption{Acceleration parameter $-q$ for the exponential potential 
$e^{\lambda \varphi}$ (solid) vs.\ $\Lambda$CDM (dotted).  The dark
and light yellow lines indicate the 1$\sigma$ and 2$\sigma$ bounds,
respectively, on $z_t$.
$\lambda$ = 1 (red), $\sqrt{2}$ (cyan), and $\sqrt{3}$ (blue)
from top to bottom at the left.}
\label{fig-exp-acc}
\end{figure}

For $\lambda^2 > 3$, the cosmological equations have a global
attractor with $\Omega_\phi = n/\lambda^2$, where $n = 3$ for the
matter dominated era (during which $w = 0$) or $n = 4$ for the
radiation dominated era (during which $w = 1/3$).  For $\lambda^2 <
3$, the cosmological equations have a late time attractor with
$\Omega_\phi = 1$ and $w = \lambda^2/3 - 1$.  In the simulations
presented here (see Figs.~\ref{fig-exp-phi}--\ref{fig-exp-acc} and
Table~\ref{exponential}), the scalar field is still evolving at $t_0$
toward the attractor solution, as advocated in
Refs.~\cite{Weller:2001gf,LopesFranca:2002ek,KL2}.


For $\lambda = \sqrt{2}$ and $\rho_m = 0$, $\ddot{a} \rightarrow 0$
asymptotically; if $\rho_m > 0$, the universe eventually enters a
future epoch of deceleration.  In either case, there is no event
horizon.  For $\lambda < \sqrt{2}$, the universe enters a period of
eternal acceleration with an event horizon.  For $\lambda > \sqrt{2}$,
the universe eventually decelerates and there is no event horizon.

The $\Lambda$CDM cosmology is approached for $\lambda \le 1/\sqrt{3}$.
Significant acceleration occurs only for $\lambda \<~ \sqrt{3}$.  For
$\lambda = \sqrt{3}$, $w_0$ is much too high; setting $\Omega_{\phi0}
= 0.66$ still results in $w_0 = -0.54$.
We conclude that $\lambda \le \sqrt{2}$ in the exponential potential
for a viable present-day cosmology.

\subsection{Stable de Sitter Cosh Potential}

The $\cosh(\sqrt{2} \varphi)$ potential exemplifies $N = 2$
supergravity with a future de Sitter space~\cite{Fre,Kallosh}, with
$m_\phi^2 = 6 H_\Lambda^2$.

\begin{table}[htb]
\center{
\begin{tabular}{|l l l l l|} \hline 
$\varphi_i$ & $\overline{w}_0$ & $z_t$ & $w_0$ & $w_1$ \\ \hline
$0.1$ & $-0.998$ & 0.67 & $-0.997$ & $0.001$ \\ 
$0.5$ & $-0.96$ & 0.68 & $-0.94$ & $0.005$ \\ 
$1$ & $-0.89$ & 0.72 & $-0.81$ & $-0.08$ \\ 
$2$ & $-0.84$ & 0.75 & $-0.69$ & $-0.30$ \\ 
\hline
\end{tabular}
}
\caption{Parameters for the potential $\cosh(\sqrt{2} \varphi)$.}
\label{cosh}
\end{table}

\begin{figure}[htbp]
\center{
\scalebox{1.1}{\includegraphics{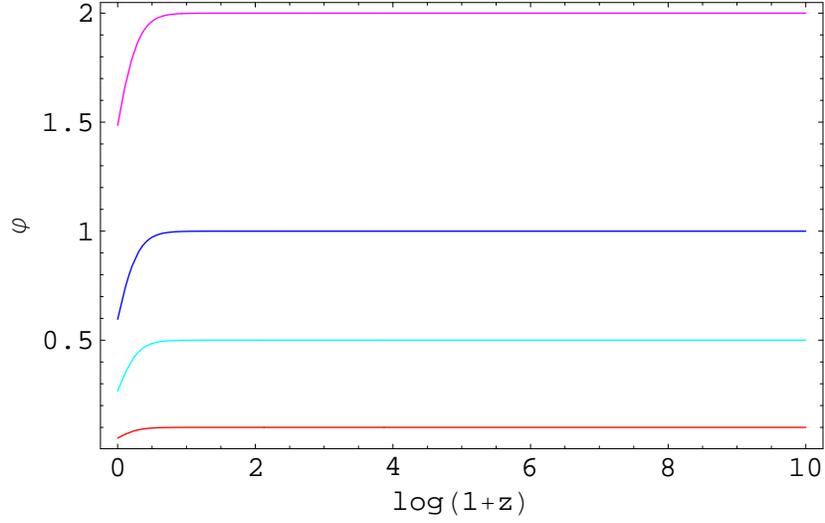}}
}
\caption{$\varphi(\tau)$ for the potential $\cosh(\sqrt{2} \varphi)$.
$\varphi_i$ = 0.1 (red), 0.5 (cyan), 1 (blue), and 2 (magenta).}
\label{fig-cosh-phi}
\end{figure}

\begin{figure}[htbp]
\center{
\scalebox{1.1}{\includegraphics{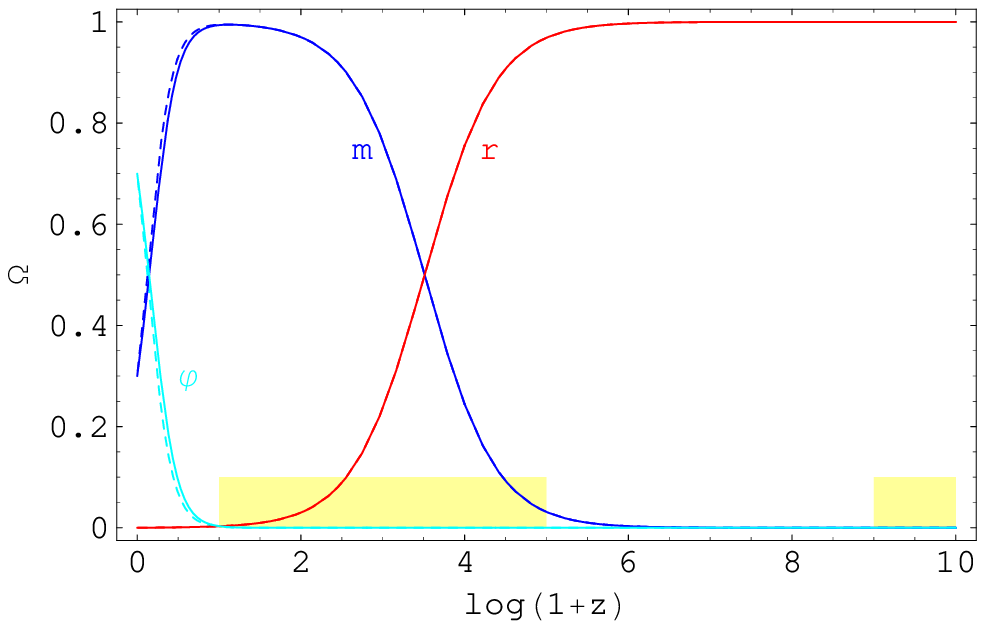}}
}
\caption{$\Omega$ for the potential $\cosh(\sqrt{2} \varphi)$,
$\varphi_i$ = 1 (solid) vs.\ $\Lambda$CDM (dotted).}
\label{fig-cosh-Omega}
\end{figure}

\begin{figure}[htbp]
\center{
\scalebox{1.1}{\includegraphics{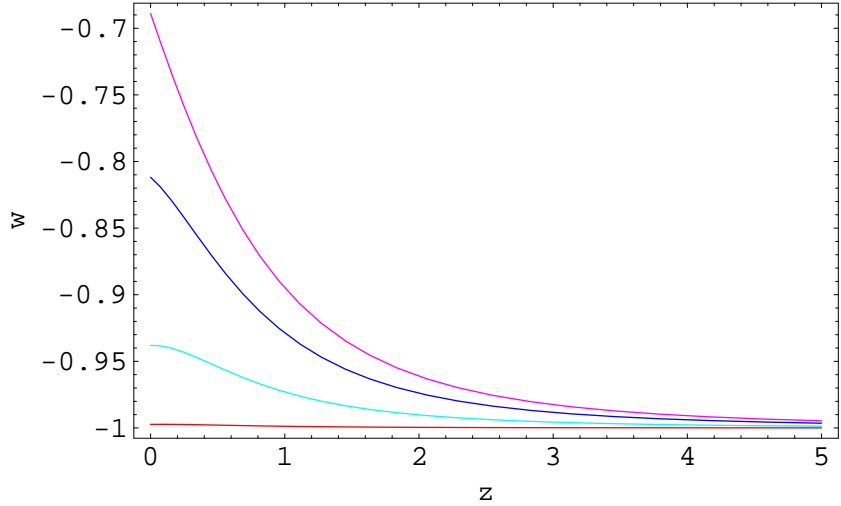}}
}
\caption{$w$ for the potential $\cosh(\sqrt{2} \varphi)$.
$\varphi_i$ = 0.1 (red), 0.5 (cyan), 1 (blue), and 2 (magenta)
from bottom to top.}
\label{fig-cosh-w}
\end{figure}

\begin{figure}[htbp]
\center{
\scalebox{1.1}{\includegraphics{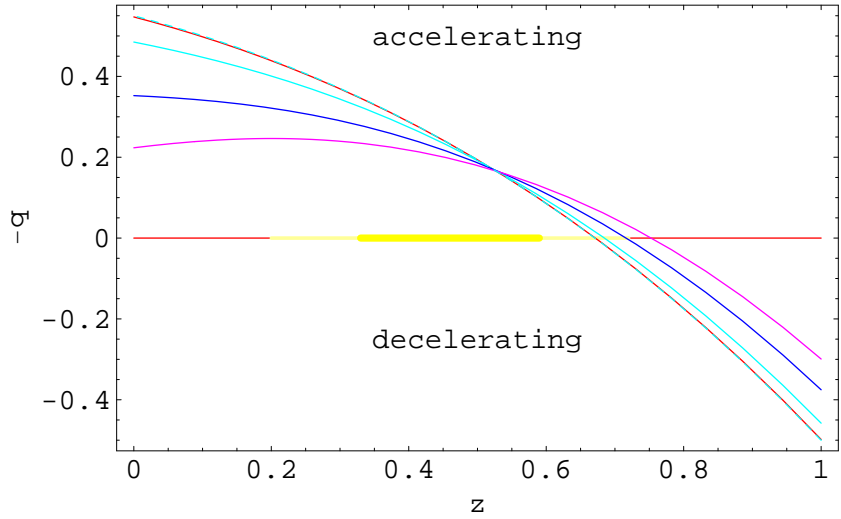}}
}
\caption{Acceleration parameter $-q$ for the potential 
$\cosh(\sqrt{2} \varphi)$ (solid) vs.\ $\Lambda$CDM (dotted).
$\varphi_i$ = 0.1 (red), 0.5 (cyan), 1 (blue), and 2 (magenta)
from top to bottom at the left.}
\label{fig-cosh-acc}
\end{figure}

Results for the cosh potential are presented in
Figs.~\ref{fig-cosh-phi}--\ref{fig-cosh-acc} and Table~\ref{cosh}.
Near $t_0$, $\varphi$ is evolving toward the minimum of the potential.
The $\Lambda$CDM model is approached as $\varphi_i \rightarrow 0$.
For $\varphi_i \ge 2$, the $\cosh(\sqrt{2} \varphi)$ results are very
nearly the same as for $e^{\sqrt{2} \varphi}$.

\subsection{Unstable de Sitter Cosh Potential}

The $2 - \cosh(\sqrt{2} \varphi)$ potential is derived from
M-theory/$N = 8$ supergravity~\cite{Hull:1984rt}, with $m_\phi^2 = - 6
H_\Lambda^2$ at the maximum of the potential.  Near the unstable de
Sitter maximum ($\varphi_i \rightarrow 0$), the universe can mimic
$\Lambda$CDM for a very long time (on the order of or greater than
$t_0$)~\cite{KL2}.

\begin{table}[htb]
\center{
\begin{tabular}{|l l l l l|} \hline 
$\varphi_i$ & $\overline{w}_0$ & $z_t$ & $w_0$ & $w_1$ \\ \hline
$0.1$ & $-0.996$ & 0.67 & $-0.99$ & $-0.04$ \\ 
$0.2$ & $-0.98$ & 0.69 & $-0.93$ & $-0.24$ \\ 
$0.3$ & $-0.92$ & 0.77 & $-0.64$ & $-1.8$ \\ 
\hline
\end{tabular}
}
\caption{Parameters for the potential $2-\cosh(\sqrt{2} \varphi)$.}
\label{2-cosh}
\end{table}

\begin{figure}[htbp]
\center{
\scalebox{1.1}{\includegraphics{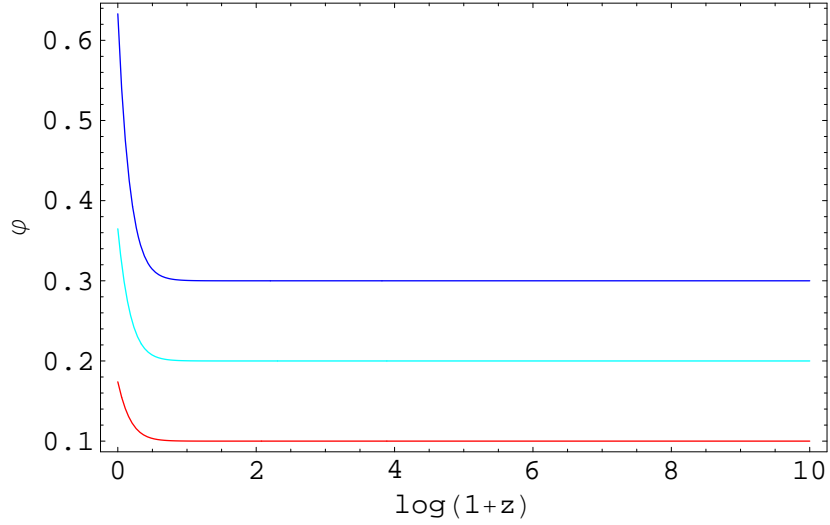}}
}
\caption{$\varphi(\tau)$ for the potential $2-\cosh(\sqrt{2} \varphi)$.
$\varphi_i$ = 0.1 (red), 0.2 (cyan), and 0.3 (blue).}
\label{fig-2-cosh-phi}
\end{figure}

\begin{figure}[htbp]
\center{
\scalebox{1.1}{\includegraphics{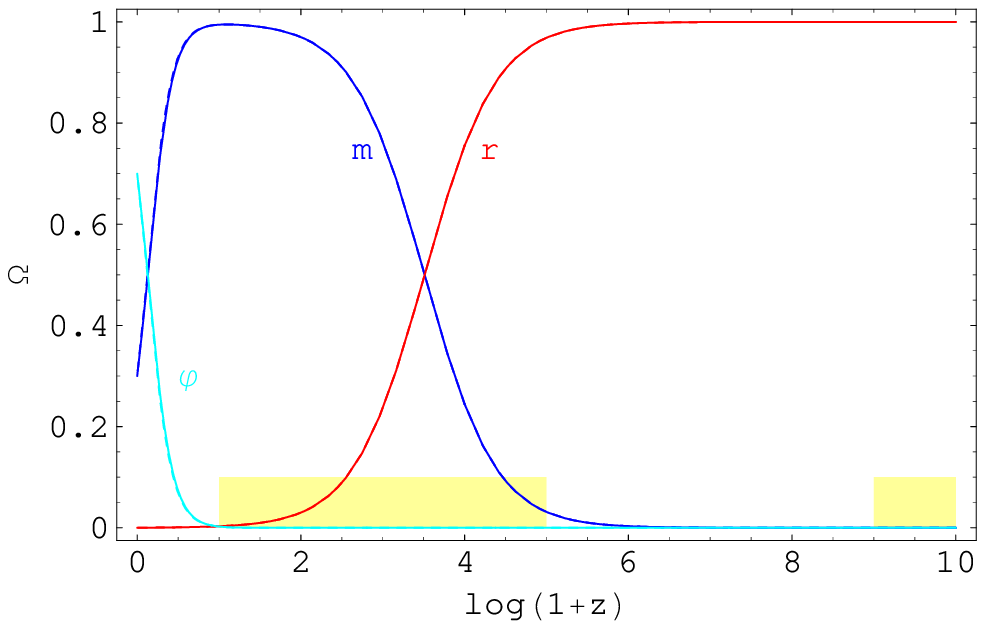}}
}
\caption{$\Omega$ for the potential $2-\cosh(\sqrt{2} \varphi)$,
$\varphi_i$ = 0.2 (solid) vs.\ $\Lambda$CDM (dotted).}
\label{fig-2-cosh-Omega}
\end{figure}

\begin{figure}[htbp]
\center{
\scalebox{1.1}{\includegraphics{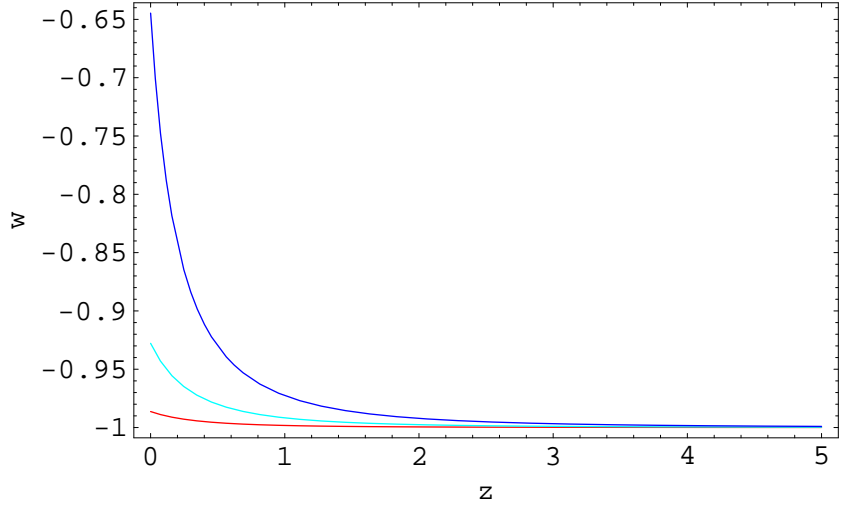}}
}
\caption{$w$ for the potential $2-\cosh(\sqrt{2} \varphi)$.
$\varphi_i$ = 0.1 (red), 0.2 (cyan), and 0.3 (blue)
from bottom to top.}
\label{fig-2-cosh-w}
\end{figure}

\begin{figure}[htbp]
\center{
\scalebox{1.1}{\includegraphics{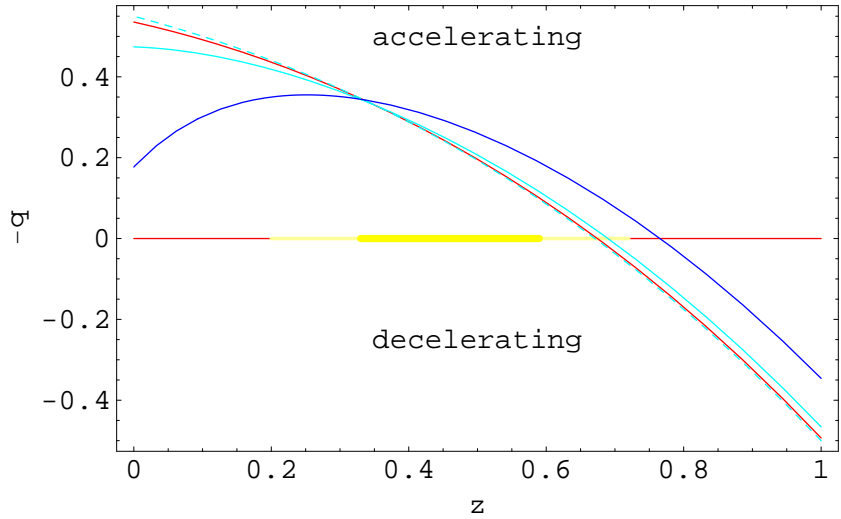}}
}
\caption{Acceleration parameter $-q$ for the potential 
$2-\cosh(\sqrt{2} \varphi)$ (solid) vs.\ $\Lambda$CDM (dotted).
$\varphi_i$ = 0.1 (red), 0.2 (cyan), and 0.3 (blue)
from top to bottom at the left.}
\label{fig-2-cosh-acc}
\end{figure}

Results for the unstable de Sitter cosh potential are presented in
Figs.~\ref{fig-2-cosh-phi}--\ref{fig-2-cosh-acc} and
Table~\ref{2-cosh}.  The scalar field is just beginning to grow
without bound at $t_0$.  For $\varphi_i \ge 0.33$, $\Omega_{\phi0}$
never reaches 0.70; for $\varphi_i \ge 0.35$, $\Omega_{\phi0}$ never
reaches 0.66.

\subsection{Axion Potential}

For the axion potentials $1+\cos(\lambda \varphi)$ and $\cos(\lambda
\varphi)$ in this and the next subsection, we can restrict our
attention to $0 \le \lambda \varphi_i \le \pi$.  We will set
$\lambda = 1$; similar results are obtained for $\lambda = \sqrt{2}$.

The axion potential $1 + \cos(\varphi)$ is based on $N = 1$
supergravity~\cite{Frieman:1995pm,Waga:2000ay}, with $m_\phi^2 = 3
H_\Lambda^2$.  As $\varphi \rightarrow \pi$, the universe evolves to
Minkowski space.

\begin{table}[htb]
\center{
\begin{tabular}{|l l l l l|} \hline 
$\varphi_i/\pi$ & $\overline{w}_0$ & $z_t$ & $w_0$ & $w_1$ \\ \hline
$0.1$ & $-0.998$ & 0.67 & $-0.995$ & $-0.01$ \\ 
$0.3$ & $-0.98$ & 0.68 & $-0.95$ & $-0.10$ \\ 
$0.5$ & $-0.90$ & 0.75 & $-0.74$ & $-0.55$ \\ 
$0.55$ & $-0.83$ & 0.82 & $-0.55$ & $-1.1$ \\ 
\hline
\end{tabular}
}
\caption{Parameters for the potential $1+\cos(\varphi)$.}
\label{1pluscos}
\end{table}

\begin{figure}[htbp]
\center{
\scalebox{1.1}{\includegraphics{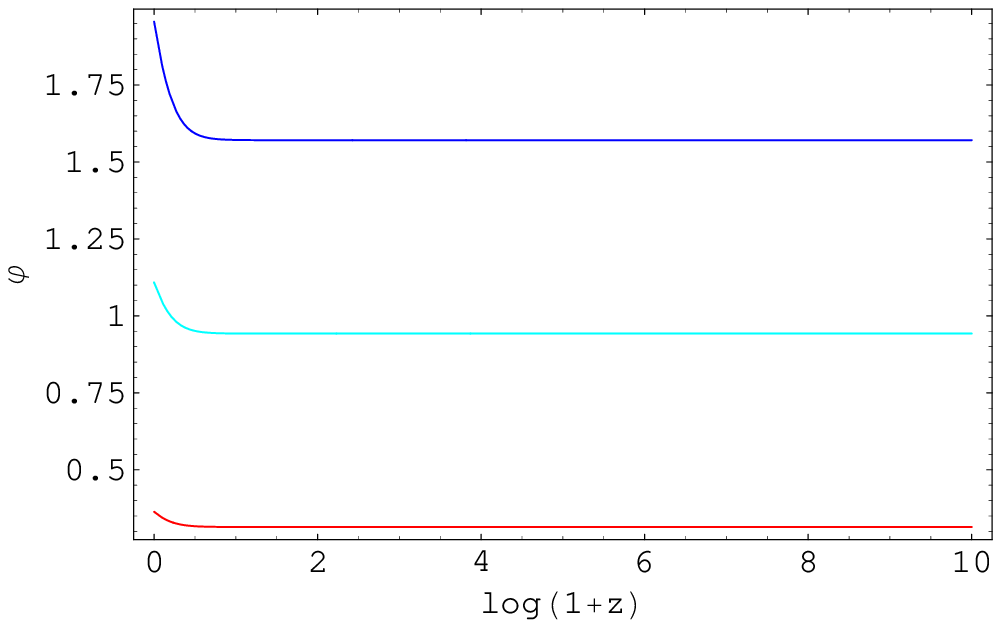}}
}
\caption{$\varphi(\tau)$ for the potential $1+\cos(\varphi)$.
$\varphi_i/\pi = 0.1$ (red), $0.3$ (cyan), and $0.5$ (blue).}
\label{fig-1pluscos-phi}
\end{figure}

\begin{figure}[htbp]
\center{
\scalebox{1.1}{\includegraphics{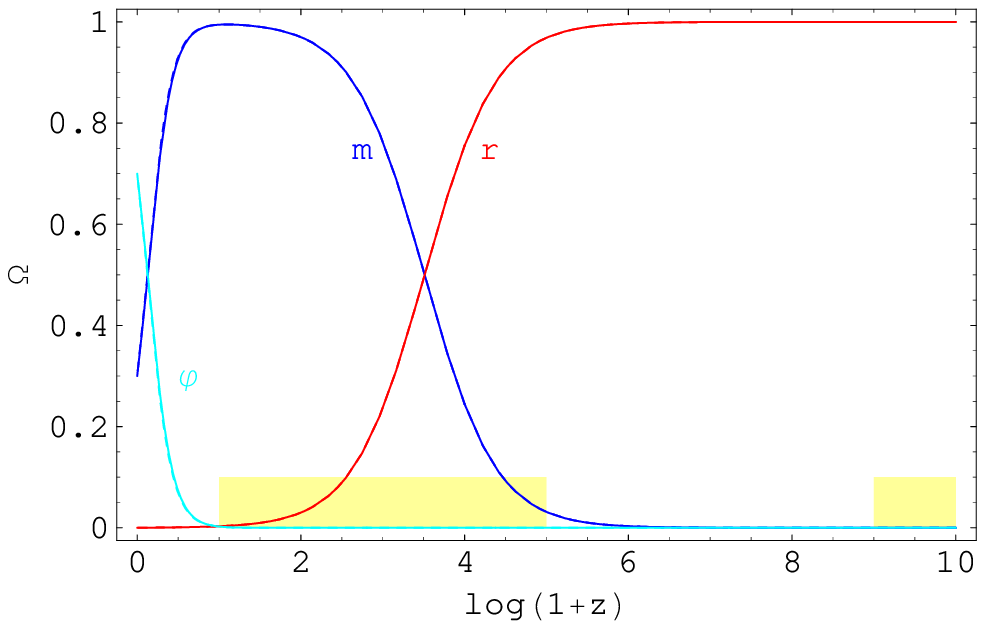}}
}
\caption{$\Omega$ for the potential $1+\cos(\varphi)$,
$\varphi_i/\pi = 0.3$ (solid) vs.\ $\Lambda$CDM (dotted).}
\label{fig-1pluscos-Omega}
\end{figure}

\begin{figure}[htbp]
\center{
\scalebox{1.1}{\includegraphics{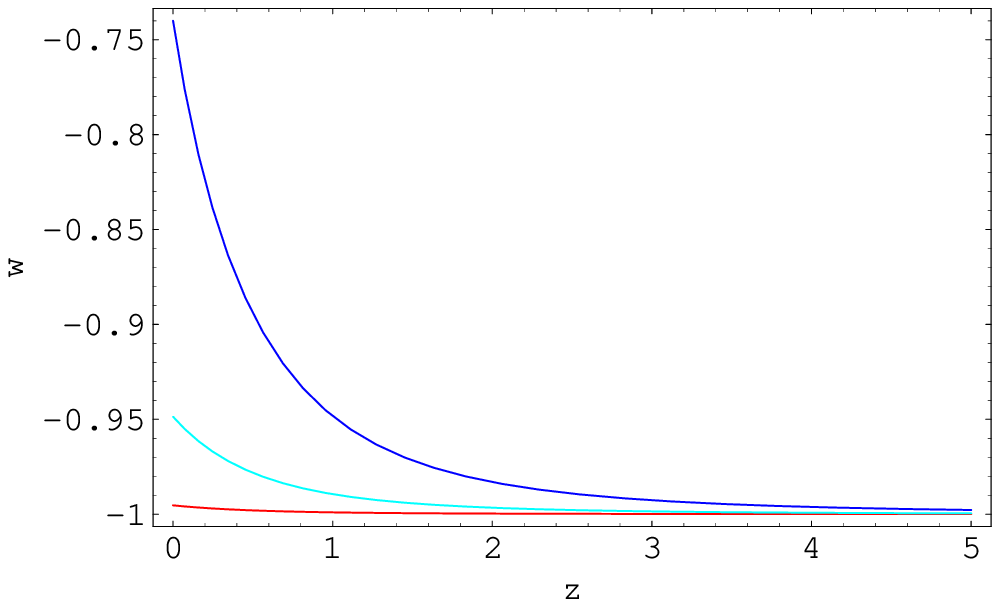}}
}
\caption{$w$ for the potential $1+\cos(\varphi)$.
$\varphi_i/\pi = 0.1$ (red), $0.3$ (cyan), and $0.5$ (blue)
from bottom to top.}
\label{fig-1pluscos-w}
\end{figure}

\begin{figure}[htbp]
\center{
\scalebox{1.1}{\includegraphics{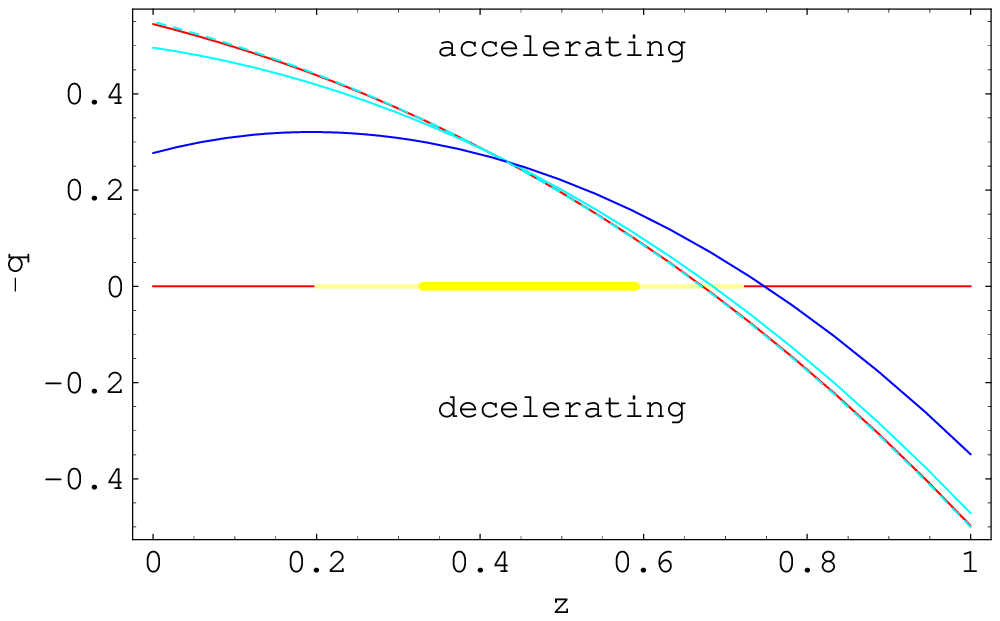}}
}
\caption{Acceleration parameter $-q$ for the potential 
$1+\cos(\varphi)$ (solid) vs.\ $\Lambda$CDM (dotted).
$\varphi_i/\pi = 0.1$ (red), $0.3$ (cyan), and $0.5$ (blue)
from top to bottom at the left.}
\label{fig-1pluscos-acc}
\end{figure}

Figures~\ref{fig-1pluscos-phi}--\ref{fig-1pluscos-acc} and
Table~\ref{1pluscos} present the results for the axion potential.  As
$\varphi_i \rightarrow 0$, a transient de Sitter universe is obtained
that mimics the $\Lambda$CDM model for a long time.  Near $t_0$,
$\varphi$ is beginning to evolve toward $\pi$.  For $\varphi_i/\pi \ge
0.59$, $\Omega_{\phi0} = 0.70$ is never attained; for $\varphi_i/\pi >
0.61$, $\Omega_{\phi0}$ never reaches 0.66.

\subsection{Unstable de Sitter Axion Potential}

The unstable de Sitter axion potential $\cos(\varphi)$ is based on
M/string theory reduced to an effective $N = 1$ supergravity
theory~\cite{Choi:1999xn}, with $m_\phi^2 = -3 H_\Lambda^2$ at the
maximum of $V$.

\begin{table}[htb]
\center{
\begin{tabular}{|l l l l l|} \hline 
$\varphi_i/\pi$ & $\overline{w}_0$ & $z_t$ & $w_0$ & $w_1$ \\ \hline
$0.05$ & $-0.998$ & 0.67 & $-0.99$ & $-0.01$ \\ 
$0.10$ & $-0.99$ & 0.68 & $-0.98$ & $-0.06$ \\ 
$0.15$ & $-0.98$ & 0.69 & $-0.93$ & $-0.16$ \\ 
$0.20$ & $-0.94$ & 0.72 & $-0.83$ & $-0.47$ \\ 
$0.25$ & $-0.78$ & 0.94 & $-0.16$ & $-4.4$ \\ 
\hline
\end{tabular}
}
\caption{Parameters for the potential $\cos(\varphi)$.}
\label{cos}
\end{table}

\begin{figure}[htbp]
\center{
\scalebox{1.1}{\includegraphics{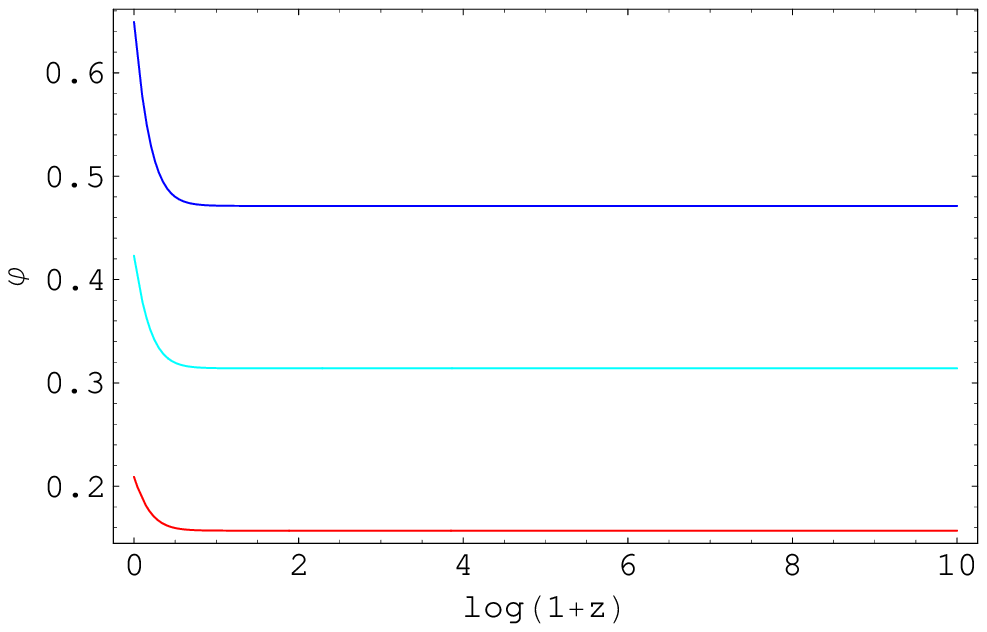}}
}
\caption{$\varphi(\tau)$ for the potential $\cos(\varphi)$.
$\varphi_i/\pi = 0.05$ (red), $0.1$ (cyan), and $0.15$ (blue).}
\label{fig-cos-phi}
\end{figure}

\begin{figure}[htbp]
\center{
\scalebox{1.1}{\includegraphics{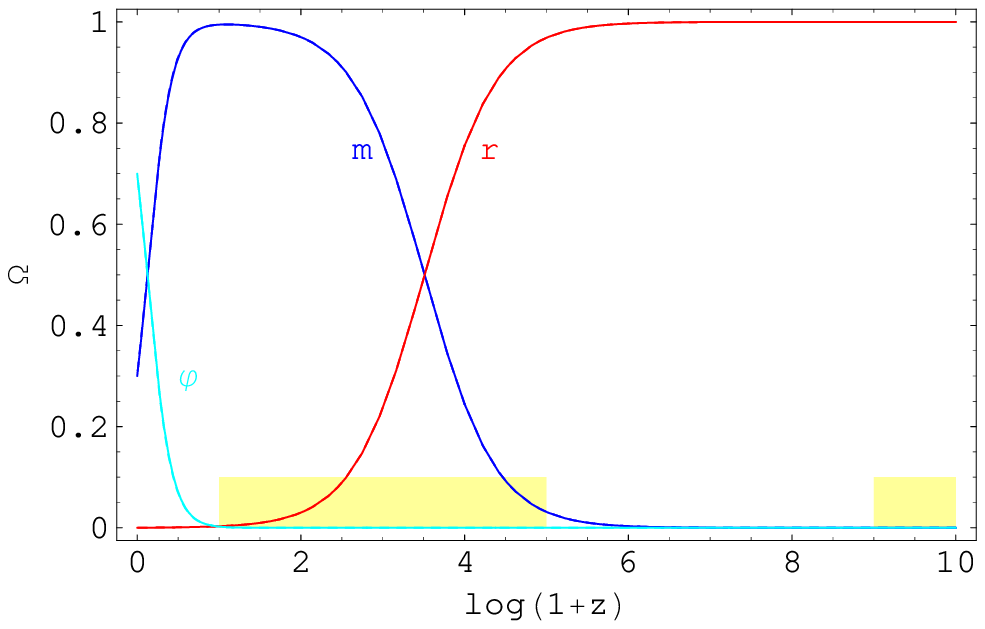}}
}
\caption{$\Omega$ for the potential $\cos(\varphi)$,
$\varphi_i/\pi$ = 0.1 (solid) vs.\ $\Lambda$CDM (dotted).}
\label{fig-cos-Omega}
\end{figure}

\begin{figure}[htbp]
\center{
\scalebox{1.1}{\includegraphics{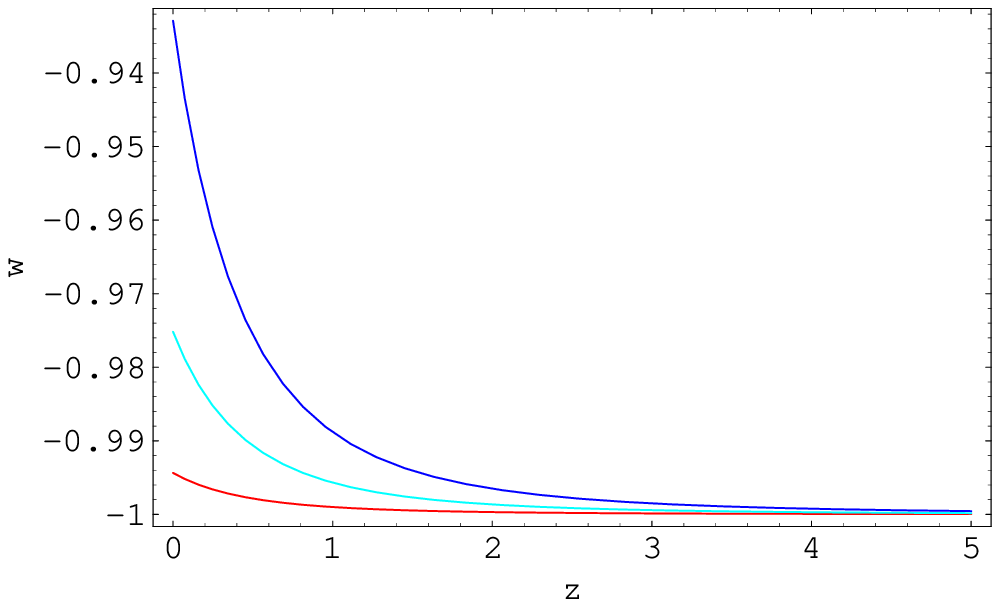}}
}
\caption{$w$ for the potential $\cos(\varphi)$.
$\varphi_i/\pi = 0.05$ (red), $0.1$ (cyan), and $0.15$ (blue)
from bottom to top.}
\label{fig-cos-w}
\end{figure}

\begin{figure}[htbp]
\center{
\scalebox{1.1}{\includegraphics{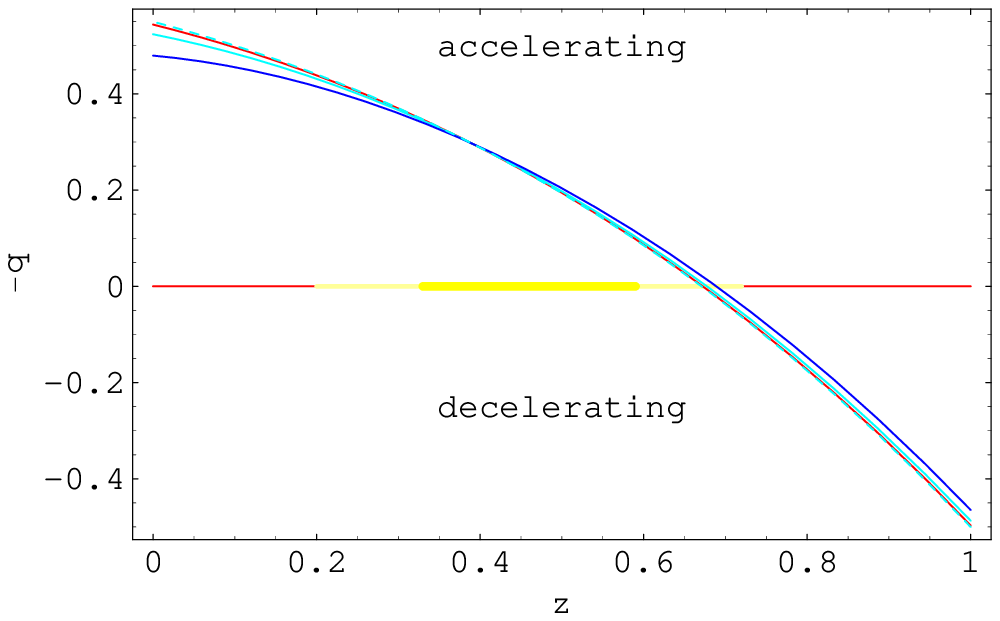}}
}
\caption{Acceleration parameter $-q$ for the potential 
$\cos(\varphi)$ (solid) vs.\ $\Lambda$CDM (dotted).
$\varphi_i/\pi = 0.05$ (red), $0.1$ (cyan), and $0.15$ (blue)
from top to bottom at the left.}
\label{fig-cos-acc}
\end{figure}

Results for the unstable axion potential are presented in
Figs.~\ref{fig-cos-phi}--\ref{fig-cos-acc} and Table~\ref{cos}.  An
unstable de Sitter universe that mimics $\Lambda$CDM for a long
time~\cite{KL2} is obtained as $\varphi_i \rightarrow 0$.  Near $t_0$,
$\varphi$ is beginning to evolve toward $\pi$.  For $\varphi_i/\pi =
0.25$, there is a transition back to deceleration at $z_t = 0.12$.

\subsection{Pol\'onyi Potential}

The Pol\'onyi potential $\left[ \left( 1 + \frac{\varphi}{\sqrt{2}}
\left(\frac{\varphi}{\sqrt{2}} + \beta\right)\right)^2 - 3
\left(\frac{\varphi}{\sqrt{2}} + \beta\right)^2 \right]
e^{\varphi^2/2}$ is derived from $N = 1$ supergravity~\cite{Polonyi}
(for a review, see Ref.~\cite{Nilles:1983ge}).  The potential is
invariant under the transformation $\varphi \rightarrow -\varphi$,
$\beta \rightarrow -\beta$.

\begin{table}[htbp]
\center{
\begin{tabular}{|l l l l l|} \hline 
$\varphi_i$ & $\overline{w}_0$ & $z_t$ & $w_0$ & $w_1$ \\ \hline
$0.05$ & $-0.85$ & 0.81 & $-0.54$ & $-1.7$ \\ 
$0$ & $-0.89$ & 0.76 & $-0.68$ & $-0.96$ \\ 
$-0.5$ & $-0.96$ & 0.69 & $-0.91$ & $-0.13$ \\ 
$-1.0$ & $-0.92$ & 0.70 & $-0.87$ & $-0.07$ \\ 
$-1.5$ & $-0.74$ & 0.57 & $-0.74$ & $0.16$ \\ 
$-1.6$ & $-0.69$ & 0.49 & $-0.73$ & $0.21$ \\ 
$-1.7$ & $-0.64$ & 0.43 & $-0.72$ & $0.26$ \\ 
$-1.8$ & $-0.59$ & 0.39 & $-0.71$ & $0.30$ \\ 
$-1.9$ & $-0.56$ & 0.36 & $-0.71$ & $0.32$ \\ 
$-2.0$ & $-0.53$ & 0.36 & $-0.72$ & $0.32$ \\ 
$-2.5$ & $-0.53$ & 0.42 & $-0.76$ & $0.25$ \\ 
\hline
\end{tabular}
}
\caption{Parameters for the Pol\'onyi potential with $\beta = 2 - \sqrt{3}$.}
\label{Pol}
\end{table}
\begin{table}[htbp]
\center{
\begin{tabular}{|l l l l l|} \hline 
$\varphi_i$ & $\overline{w}_0$ & $z_t$ & $w_0$ & $w_1$ \\ \hline
$-1.5$ & $-0.74$ & 0.55 & $-0.77$ & $0.21$ \\ 
$-1.6$ & $-0.69$ & 0.49 & $-0.76$ & $0.27$ \\ 
$-1.7$ & $-0.64$ & 0.43 & $-0.75$ & $0.32$ \\ 
$-1.8$ & $-0.60$ & 0.40 & $-0.75$ & $0.35$ \\ 
$-1.9$ & $-0.56$ & 0.39 & $-0.76$ & $0.36$ \\ 
$-2.0$ & $-0.54$ & 0.38 & $-0.76$ & $0.36$ \\ 
\hline
\end{tabular}
}
\caption{Parameters for the Pol\'onyi potential with $\beta = 0.2$.}
\label{Pol.2}
\end{table}
\begin{table}[htbp]
\center{
\begin{tabular}{|l l l l l|} \hline 
$\varphi_i$ & $\overline{w}_0$ & $z_t$ & $w_0$ & $w_1$ \\ \hline
$-1.5$ & $-0.73$ & 0.58 & $-0.68$ & $0.02$ \\ 
$-1.6$ & $-0.68$ & 0.49 & $-0.66$ & $0.07$ \\ 
$-1.7$ & $-0.63$ & 0.40 & $-0.64$ & $0.12$ \\ 
$-1.8$ & $-0.58$ & 0.35 & $-0.63$ & $0.15$ \\ 
$-1.9$ & $-0.54$ & 0.32 & $-0.63$ & $0.17$ \\ 
$-2.0$ & $-0.51$ & 0.30 & $-0.63$ & $0.18$ \\ 
\hline
\end{tabular}
}
\caption{Parameters for the Pol\'onyi potential with $\beta = 0.4$.}
\label{Pol.4}
\end{table}

\begin{figure}[htbp]
\center{
\scalebox{1.1}{\includegraphics{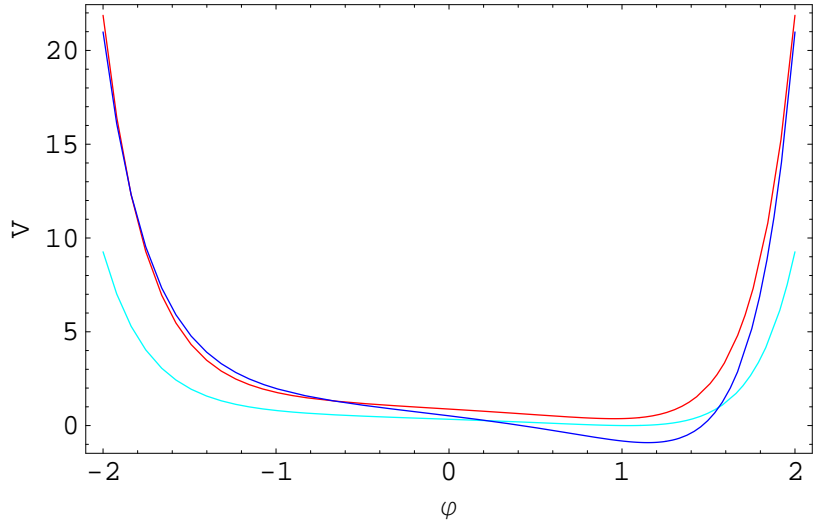}}
}
\caption{Dimensionless Pol\'onyi potential for 
$\beta = 2-\sqrt{3}$ (cyan, $V_{min} = 0$), 
0.2 (red, $V_{min} > 0$), and 0.4 (blue, $V_{min} < 0$).}
\label{fig-Pol}
\end{figure}

\begin{figure}[htbp]
\center{
\scalebox{1.1}{\includegraphics{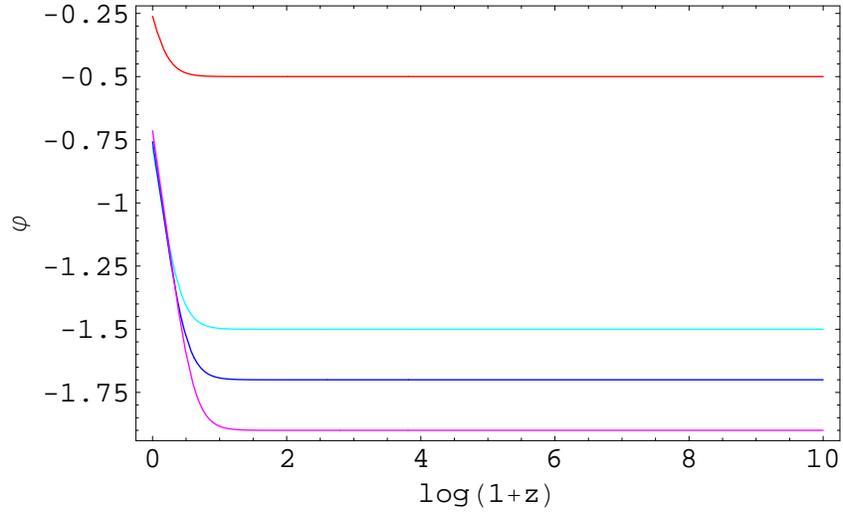}}
}
\caption{$\varphi(\tau)$ for the Pol\'onyi potential.
$\varphi_i = -0.5$ (red), $-1.5$ (cyan), $-1.7$ (blue), and $-1.9$ (magenta).}
\label{fig-Pol-phi}
\end{figure}

\begin{figure}[htbp]
\center{
\scalebox{1.1}{\includegraphics{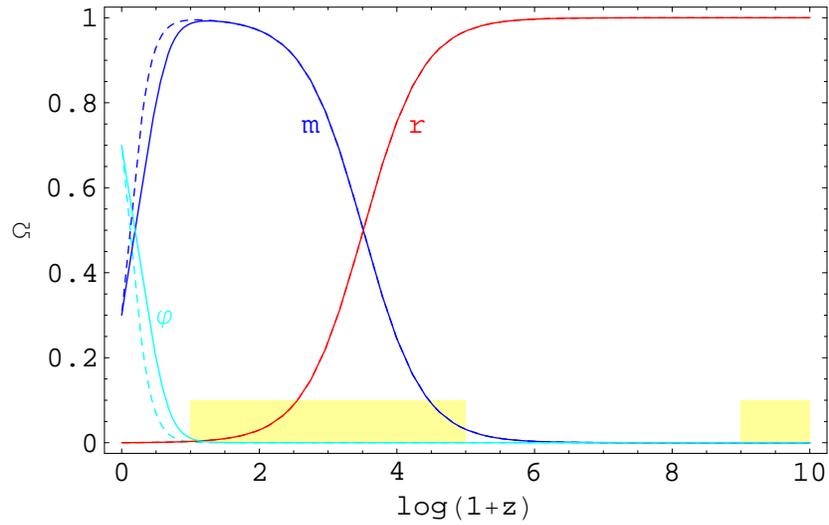}}
}
\caption{$\Omega$ for the Pol\'onyi potential.
$\varphi_i = -1.7$ (solid) vs.\ $\Lambda$CDM (dotted).}
\label{fig-Pol-Omega}
\end{figure}

\begin{figure}[htbp]
\center{
\scalebox{1.1}{\includegraphics{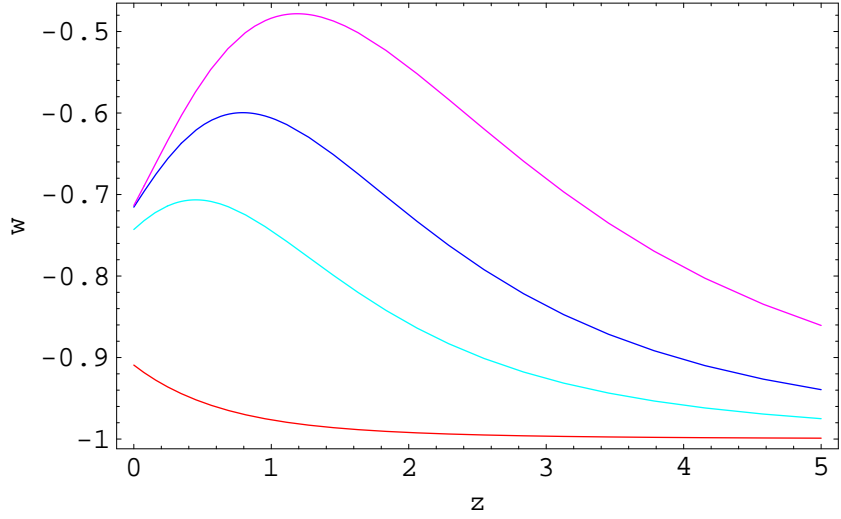}}
}
\caption{$w$ for the Pol\'onyi potential.
$\varphi_i = -0.5$ (red), $-1.5$ (cyan), $-1.7$ (blue), and $-1.9$ (magenta)
from bottom to top.}
\label{fig-Pol-w}
\end{figure}

\begin{figure}[htbp]
\center{
\scalebox{1.1}{\includegraphics{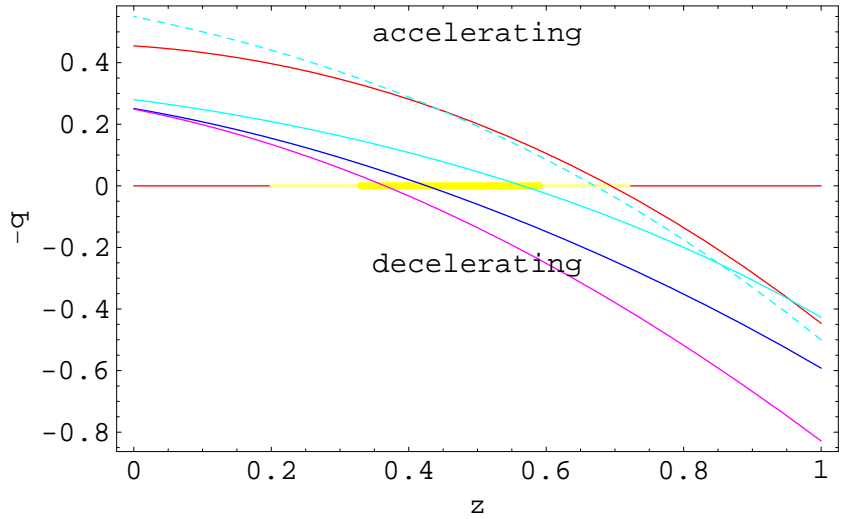}}
}
\caption{Acceleration parameter $-q$ for the Pol\'onyi potential (solid)
vs.\ $\Lambda$CDM (dotted).
$\varphi_i = -0.5$ (red), $-1.5$ (cyan), $-1.7$ (blue), and $-1.9$ (magenta)
from top to bottom at the left.}
\label{fig-Pol-acc}
\end{figure}

Following Ref.~\cite{KL2}, we will take $m_\phi^2 \sim
\rho_\Lambda/M_P^2$ and set $\beta = 2-\sqrt{3}$, 0.2, and 0.4, for
which the universe asymptotically evolves to Minkowski space, de
Sitter space, or a collapse respectively (see Fig.~\ref{fig-Pol}).
Figures~\ref{fig-Pol-phi}--\ref{fig-Pol-acc} and Table~\ref{Pol} have
$\beta = 2-\sqrt{3}$.  For this value of $\beta$, $\Omega_{\phi0} =
0.70$ is not reached for $\varphi_i \ge 0.09$.  $\Omega_\phi$ begins
to violate the LSS bound as $\varphi_i$ goes below $-2.5$.  The
$\Lambda$CDM model is approximated for $\varphi_i \approx -0.5$.  At
$t_0$, $\varphi$ is beginning to evolve toward the location $\varphi =
\sqrt{2} (\sqrt{3}-1)$ of the minimum of the potential.  For $-2.0 \<~
\varphi_i \<~ -1.5$, $z_t \approx 0.5$ and at least $w_0$ and $w_1$
satisfy the observational bounds.

Tables~\ref{Pol.2} and~\ref{Pol.4} demonstrate that a transition
redshift $z_t \approx$ 0.4--0.5 is not an accident due to the
particular choice $\beta = 2-\sqrt{3}$.

\subsection{SUGRA Potential}

The SUGRA potential $e^{\varphi^2/2}/\varphi^\alpha$ is derived from
$N = 1$ supergravity~\cite{Brax:1999gp,B-M,Brax:2001ah,Copeland:2000vh}.  
The minimum of the potential occurs at $\varphi = \sqrt{\alpha}$, and
$m_\phi^2 = 6 H_\Lambda^2$.  We will take $\alpha = 4$, which has the
interesting property that the minimum of the potential $V_{min} \sim
M^8/M_P^4 \sim \rho_\Lambda$ for $M \sim M_{weak} \sim$ 1
TeV~\cite{Brax:2001ah}.

\begin{table}[htb]
\center{
\begin{tabular}{|l l l l l|} \hline 
$\varphi_i$ & $\overline{w}_0$ & $z_t$ & $w_0$ & $w_1$ \\ \hline
$10^{-6}$ & $-0.68$ & 0.50 & $-0.86$ & $0.35$ \\ 
$0.1$ & $-0.68$ & 0.50 & $-0.86$ & $0.35$ \\ 
$0.5$ & $-0.67$ & 0.50 & $-0.86$ & $0.36$ \\ 
$1$ & $-0.74$ & 0.53 & $-0.82$ & $0.36$ \\ 
$1.5$ & $-0.94$ & 0.68 & $-0.93$ & $0.06$ \\ 
$1.9$ & $-0.998$ & 0.67 & $-0.997$ & $0.001$ \\ 
$2.1$ & $-0.998$ & 0.67 & $-0.997$ & $0.001$ \\ 
$2.5$ & $-0.96$ & 0.68 & $-0.94$ & $0.01$ \\ 
$3$ & $-0.85$ & 0.69 & $-0.79$ & $0.07$ \\ 
$3.5$ & $-0.65$ & 0.39 & $-0.63$ & $0.26$ \\ 
$4$ & $-0.44$ & 0.14 & $-0.57$ & $0.53$ \\ 
\hline
\end{tabular}
}
\caption{Parameters for the SUGRA potential.}
\label{SUGRA}
\end{table}

\begin{figure}[htbp]
\center{
\scalebox{1.1}{\includegraphics{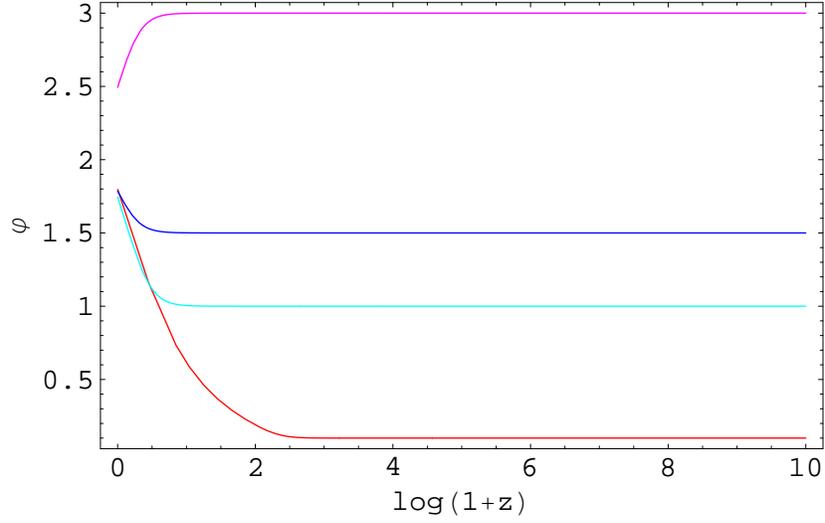}}
}
\caption{$\varphi(\tau)$ for the SUGRA potential.
$\varphi_i = 0.1$ (red), 1 (cyan), 1.5 (blue), and 3 (magenta).}
\label{fig-SUGRA-phi}
\end{figure}

\begin{figure}[htbp]
\center{
\scalebox{1.1}{\includegraphics{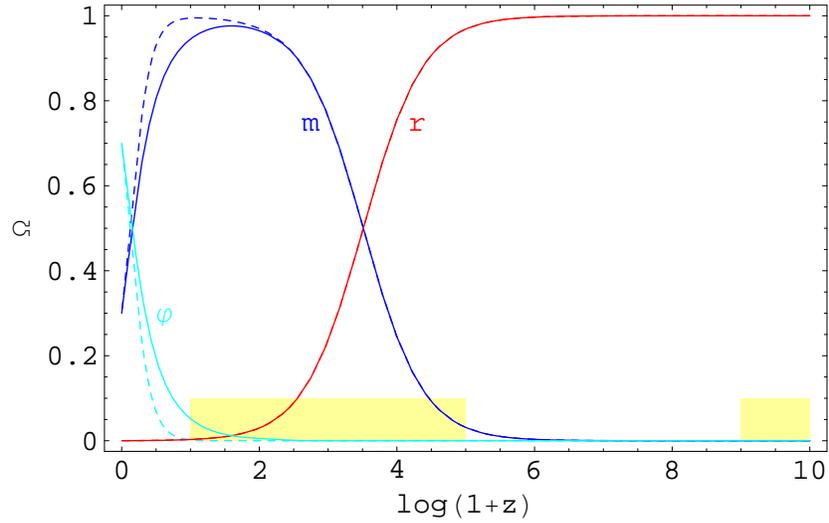}}
}
\caption{$\Omega$ for the SUGRA potential.
$\varphi_i = 0.1$ (solid) vs.\ $\Lambda$CDM (dotted).}
\label{fig-SUGRA-Omega}
\end{figure}

\begin{figure}[htbp]
\center{
\scalebox{1.1}{\includegraphics{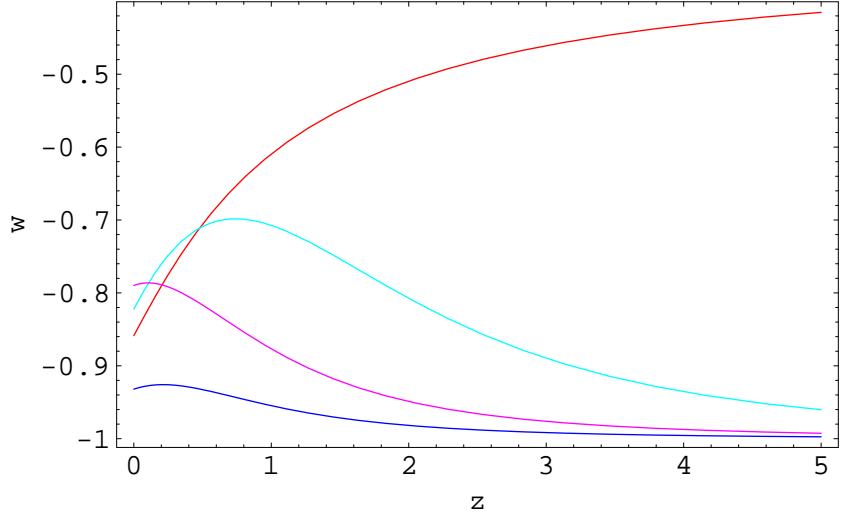}}
}
\caption{$w$ for the SUGRA potential.
$\varphi_i = 0.1$ (red, top), 1 (cyan, second from top), 
1.5 (blue, bottom), and 3 (magenta, third from top) at the right.}
\label{fig-SUGRA-w}
\end{figure}

\begin{figure}[htbp]
\center{
\scalebox{1.1}{\includegraphics{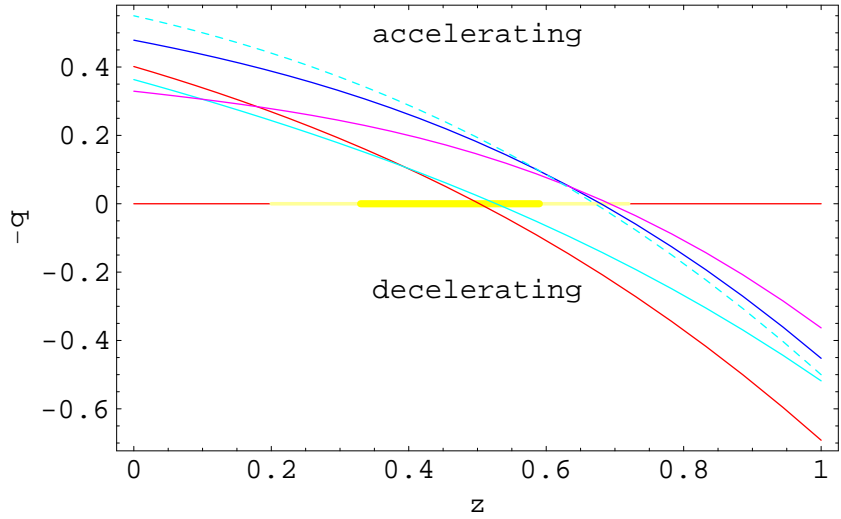}}
}
\caption{Acceleration parameter $-q$ for the SUGRA potential (solid)
vs.\ $\Lambda$CDM (dotted).
$\varphi_i = 0.1$ (red), 1 (cyan), 1.5 (blue), and 3 (magenta)
from bottom to top at the right.}
\label{fig-SUGRA-acc}
\end{figure}

Results for the SUGRA potential are presented in
Figs.~\ref{fig-SUGRA-phi}--\ref{fig-SUGRA-acc} and Table~\ref{SUGRA}.
At present $\varphi$ is evolving toward the location of the minimum of
the potential.  For $\varphi_i$ near the minimum of $V$ at $\varphi =
2$, the SUGRA potential cosmology approaches $\Lambda$CDM.  For
$\varphi_i \ge 4$, $\overline{w}_0$ and $w_0$ are much too high.

The transition redshift $z_t \approx 0.5$ for $0 < \varphi_i \<~ 1$.
For $0 < \varphi_i \le 0.5$, asymptotic values $\overline{w}_0 =
-0.68$, $z_t = 0.50$, $w_0 = -0.86$, and $w_1 = 0.35$ are obtained,
which makes these SUGRA model values robust.  These asymptotic values
are in excellent agreement with the observed central values.  (There
is also a very small interval $\varphi_i = 3.3$--$3.55$ which yields
$z_t$ = 0.33--0.59.)

\section{Conclusion}

All seven potentials can closely mimic the $\Lambda$CDM model at low
redshifts, but only the SUGRA and Pol\'onyi potentials can realize a
transition redshift of $z_t \approx 0.5$ for $\Omega_{\Lambda0} =
0.70$.  The other five models predict $z_t \ge 0.67$.

The SN Ia central value $z_t \approx 0.5$ can naturally be explained
either by the SUGRA potential with $0 < \varphi_i \<~ 1$ or by the
Pol\'onyi potential with $-2.0 \<~ \varphi_i \<~ -1.5$.  For just the
solutions with $z_t \approx 0.5$, (i) $\Omega_\phi$ becomes
significant noticeably earlier than $\Omega_\Lambda$ for $\Lambda$CDM
and (ii) either $w$ has a maximum near $z = 1$ or $w$ evolves rapidly
between $z = 5$ and the present (SUGRA $0 < \varphi_i \le 0.5$).  The
SUGRA range of initial values does not involve fine tuning, and has
the advantage of also offering a explanation (when $\alpha = 4$) of
the parametric relationship $\rho_\Lambda \sim M_{weak}^8/M_P^4$.

The low-$z$ ($0 \le z \le 5$) data on $z_t$, $\overline{w}_0$, $w_0$,
and $w_1$, although clearly capable of ruling out a cosmological
constant, cannot easily distinguish between the stable and unstable de
Sitter cases for the cosh potentials, between the two axion
potentials, or among the three different Pol\'onyi potential cases.
There is no clear distinguishing signal like the sign of $w_1$.
However, knowledge of $w(z)$ for $0 \le z \le 5$ does hold out the
prospect---if $\Omega_{\Lambda0}$ is actually due to quintessence---of
determining which quintessence potential nature may have chosen.


\clearpage

\end{document}